\documentclass[conference,letterpaper]{IEEEtran}

\usepackage[english]{babel}
\usepackage[utf8]{inputenc}
\usepackage[T1]{fontenc}
\usepackage{amsmath,amsfonts}
\usepackage{algorithm,algorithmic}
\usepackage{graphicx}
\usepackage{epstopdf}
\usepackage{xcolor}
\usepackage{booktabs}
\usepackage{enumitem}
\usepackage{cite}
\usepackage{url}
\usepackage{float}
\usepackage{hyperref}

\definecolor{backgroundbox}{rgb}{0,0,0}
\def\BibTeX{{\rm B\kern-.05em{\sc i\kern-.025em b}\kern-.08em
    T\kern-.1667em\lower.7ex\hbox{E}\kern-.125emX}}

\newcommand{\HH}[1]{#1}
\newcommand{\IN}[1]{#1}

\begin{document}

\title{PRISM: RL-Based Rate Control for Low Latency and High Data Rate in Mobile Edge Networks}

\author{\IEEEauthorblockN{Hamid Hassani, Staji Wilson, Kishor C. Joshi, Sonia M. Heemstra, Ignas G. Niemegeers, George Exarchakos}
\IEEEauthorblockA{Eindhoven University of Technology, The Netherlands\\
{\tt \{h.hassani,s.wilson,k.c.joshi,sheemstradegroot,i.g.m.niemegeers,g.exarchakos\}@tue.nl}}
}

\maketitle

\begin{abstract}
Emerging delay-sensitive interactive services, such as extended reality and remote operation, demand advanced congestion control algorithms (CCAs) capable of maintaining high data rates while keeping latency low in mobile networks.
Existing CCAs typically rely on congestion signals such as packet loss, delay, or explicit congestion notification (ECN) marks, which can be delayed, imprecise, or require extensive upgrades to the network infrastructure. 
To address these limitations, physical resource block (PRB) utilization has recently been proposed as a reliable early congestion signal in mobile networks~\cite{Hassani25a}. 
This paper introduces PRISM, a reinforcement-learning-based CCA 
that uses peak PRB utilization to regulate the send rate
under challenging mobility and fading conditions. 
NS3 simulation results show that, in a mobility scenario where two PRISM flows coexist with two TCP Cubic flows and one TCP BBR flow, PRISM maintains a per-packet mean one-way delay (OWD) of $6.3$ms and a $p_{99}$ OWD of $44.6$ms, with negligible ($0.02\%$) packet loss, over a $20$min high-contention interval. 
However, over the same interval, the coexisting TCP flows experience $60\times$ higher mean OWDs of $371$-$390$ms and $p_{99}$ OWDs of $563$-$578$ms, with $0.1\%$ packet loss. 
PRISM achieves a $31\times$ improvement in goodput-over-latency performance, indicating a favorable tradeoff for delay-sensitive applications. 
\end{abstract}
\section{Introduction}
Enabling high data rate and low latency communication for next-generation interactive applications remains a key challenge for end-to-end congestion control algorithms (CCAs) in 5G and beyond mobile networks. 
Existing CCAs may prove insufficient for emerging applications that demand both high data rate and low delay \cite{Vleeschauwer24}. 

Using packet loss as a congestion signal, as in TCP Cubic~\cite{Ha08}, often leads to significant delays, because the send rate is reduced only after substantial queue buildup has already occurred \cite{Tanenbaum21}. 
The use of round-trip time (RTT) or one-way delay (OWD) as congestion signals is also challenging: RTT (as in TCP BBR~\cite{BBR}) is known to be an imprecise indicator of congestion \cite{Chefrour21}, while OWD is difficult to measure accurately in real-world deployments \cite{Hassani21}. 
Furthermore, explicit congestion notification (ECN) mechanisms, such as L4S \cite{RFC9330}, are often impractical due to deployment and infrastructure upgrade difficulties. 

Learning-based CCAs have been proposed to enhance adaptability under changing network conditions. 
However, recent results in \cite{Larrea26} show that such CCAs, including PCC \cite{Dong15} and PCC-Vivace \cite{Dong18}, can perform inconsistently in mobile networks, where mobility can cause rapid bandwidth fluctuations that are difficult to infer from end-to-end transport signals. 
Furthermore, reinforcement learning (RL)-based CCAs, such as Orca~\cite{Abbasloo20} and ReCoCo~\cite{Dena26}, use RL to adapt congestion-control policies to complex network dynamics. 
However, they mainly depend on generic transport-layer signals, such as throughput, delay, and packet loss, which may be insufficient to achieve both high data rate and low latency~\cite{Vleeschauwer24}. 

A recent study showed that physical resource block (PRB) utilization can serve as an effective early indicator of congestion in mobile networks, as it is correlated with queueing delay at the RLC layer of the base station (BS)~\cite{Hassani25a}. 
The proposed CCA, RAPID, regulates the send rate to maintain PRB utilization near a desired target, enabling both low latency and high data rate in private 5G edge networks. 
RAPID relies on a proportional–integral (PI) control policy with a fixed feedback gain, see Algorithm 1 in \cite{Hassani25a}. 
When wireless channel conditions vary due to, e.g., mobility and fading, this constant gain may become either overly aggressive or overly conservative, leading to unstable rate control and significant OWD spikes. 
Moreover, using only the average PRB utilization for rate control can hide short radio-resource bursts within a control interval. Such bursts can briefly build up the radio link control (RLC) queue and cause high OWD spikes~\cite{Balasingam19,Karnam26} even when the average PRB utilization remains below the target. 

To address the above challenges, we present \emph{PRISM}, a novel user-space transport solution based on proximal policy optimization (PPO) \cite{Schulman17}, in which the clipped objective stabilizes policy updates under noisy measurements and nonstationary contention. 
PRISM enables low delay, high data rate communication in challenging mobile network scenarios with user equipment (UE) mobility. 
It leverages cross-layer state features, including PRB utilization statistics and radio-quality indicators, to adjust the feedback gain at runtime. 
We implement PRISM in NS3 and evaluate its performance in the presence of UE mobility and cross-traffic. 

The main contributions of this paper are as follows:
\begin{itemize}
    \item We design and implement PRISM, a PPO-based CCA that uses cross-layer state features, including PRB utilization statistics and channel-quality information, to adapt its control gain and regulate the send rate based on peak PRB utilization in dynamic mobile environments. 
    \item NS3 simulation results under UE mobility and TCP cross-traffic show that, in a scenario where two PRISM flows coexist with two Cubic flows and one BBR flow, PRISM maintains a mean OWD of $6.3$ms and a $p_{99}$ OWD of $44.6$ms, about $60\times$ and $13\times$ lower, respectively, than the coexisting TCP flows, while achieving a $31\times$ improvement in goodput-over-latency performance. 
\end{itemize}

\HH{The remainder of the paper is organized as follows. 
We present the system model in Section~\ref{sec:system}. 
In Section~\ref{sec:prism}, we describe the design of PRISM, including its learning formulation and the training and model-selection procedure. 
Next, we discuss the performance evaluation results in Section~\ref{sec:results} 
and conclude the paper in Section~\ref{sec:conclusion}.}

\section{System Model} \label{sec:system}
\begin{figure}[tbp]
    \centering
    \includegraphics[width=0.85\columnwidth]{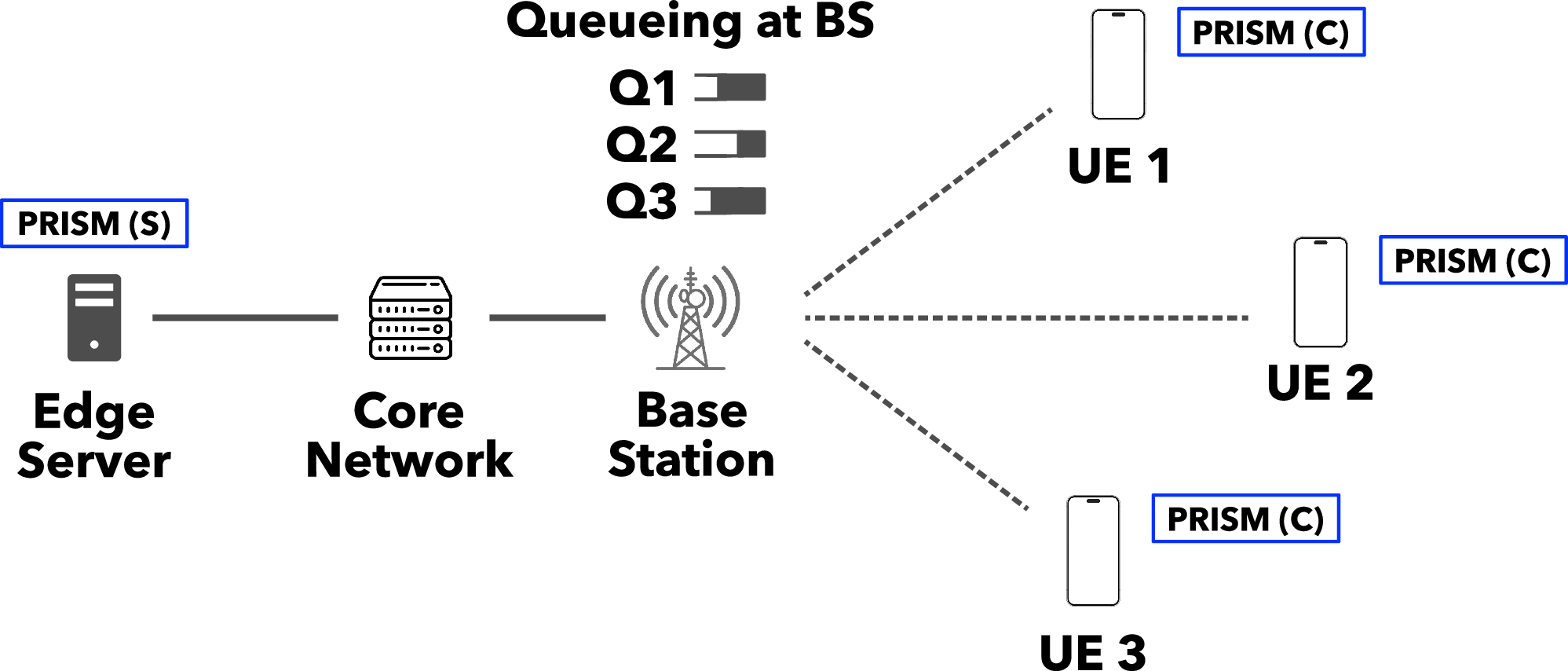}
    \caption{Edge transport architecture with the path bottleneck at the BS.}
    \label{fig:system}
\end{figure}

We consider a 5G edge architecture, as depicted in Fig.~\ref{fig:system}, where the path bottleneck lies at the RLC layer of the BS~\cite{Hassani25b}. \HH{The BS adheres to a per-data-radio-bearer (DRB) queueing policy~\cite{Irazabal24}.} 
The network is a private 5G in which all traffic flows are under the administration of the edge server (ES). 
We use the proportional-fair (PF) scheduler, 
which is widely adopted as a reference baseline in mobile network research~\cite{Huang18,Patriciello19,5gLENA}. 
The downlink traffic type is considered to be identical, e.g., best effort, so that all traffic flows towards a UE are carried by a single DRB, thus sharing the same RLC buffer at the BS. Moreover, we presume a full-buffer model, where there is always packet(s) queued in the RLC buffer of each UE at scheduling decisions~\cite{Huang18}. 
\section{PRISM Design} \label{sec:prism}
We introduce a feedback loop that adjusts the downlink send rate at the ES. 
UE $i$ collects cross-layer measurements, \HH{such as PRB utilization and transport-block (TB) retransmission ratio}, and reports this information to the ES via UDP every $\delta$ seconds. 
The ES operates over control intervals of duration $\Delta$ ($>\delta$) and uses the received reports to construct the PPO state and update the send rate of UE $i$. 

RAPID~\cite{Hassani25a} updates the send rate of UE $i$ as follows: 
\begin{equation} \label{eq:rapid}
    r^{\mathrm{RAPID}}_{i,k+1} \gets r^{\mathrm{RAPID}}_{i,k} - K \cdot \gamma \cdot \left(\mu_{\text{PRB}_{i}}(\Delta_k) - P_{\epsilon,i}\right)
\end{equation}
where $r^{\mathrm{RAPID}}_{i,k}$ is the send rate of UE~$i$, $\mu_{\mathrm{PRB}_{i}}(\Delta_k)$ denotes its average PRB utilization over $\Delta_k$, and $P_{\epsilon,i}$ is its target PRB utilization.
\HH{The parameter $K > 0$ is a constant control gain that balances the convergence rate and stability of RAPID, and $\gamma$ is a scaling factor that converts the PRB utilization error into a rate adjustment.} 
\HH{Typical design parameter values are $\Delta = 200$ms, $\delta = 50$ms, and $K = 0.5$ \cite{Hassani25a}.} 

\HH{Through NS3 simulations, we observed that using average PRB utilization, as in RAPID, can mask short bursts of high PRB utilization that arise under UE mobility, e.g., when UEs experience favorable channel conditions due to close proximity to the BS. 
These transient bursts are critical early indicators of congestion in mobile networks, as they may indicate queue buildup at the BS~\cite{Balasingam19}. 
This motivated us to explore alternative PRB-utilization statistics that better capture these transient high PRB utilization phases.} 

\HH{While one possible solution would be to reduce both the feedback interval $\delta$ and the control interval $\Delta$, this approach has important drawbacks: a smaller $\delta$ increases uplink feedback overhead, while a shorter $\Delta$ can lead to more abrupt and unstable send-rate updates. Therefore, both effects are undesirable for scalable, reliable, and stable operation.} 

PRISM preserves RAPID's framework but (i) replaces the constant feedback gain $K$ with an adaptive gain, and (ii) uses maximum PRB utilization instead of average PRB utilization. 
\HH{The PPO agent learns to select this gain based on a state formed from a short history of recent observations.} 

Let $\mathcal{N}_{i,k}$ denote the set of feedback reports delivered by UE $i$ during interval $\Delta_{k}$. 
We denote $P^{\max}_{\mathrm{PRB},i}(\Delta_k) = \max_{j \in \mathcal{N}_{i,k}} \mathrm{PRB}_{i,j}$ as the maximum PRB utilization of UE $i$ during interval $\Delta_{k}$, where $\mathrm{PRB}_{i,j}$ is the PRB utilization of UE $i$ within the interval $\delta_j$. 
If $P^{\max}_{\mathrm{PRB},i}(\Delta_k)$ is less than its target value $P_{\epsilon,i}$, the ES increases the send rate $r_i$ and decreases it if $P^{\max}_{\mathrm{PRB},i}(\Delta_k) > P_{\epsilon,i}$. 

Unlike RAPID, which uses a fixed gain $K$ in \eqref{eq:rapid}, PRISM uses a PPO-selected adaptive gain $a_{i,k}$ to determine the magnitude of the rate update for UE $i$ during the control interval $\Delta_k$. 
For rate increases, the implementation scales $a_{i,k}$ by $1/n$, where $n$ denotes the number of active downlink flows. 
Thus, ramp-up becomes more conservative under higher contention, while the PPO-selected gain is used for rate decreases. 
This feedback control logic is implemented using the pseudocode shown in Algorithm \ref{alg:ctrl}. 
\begin{algorithm}[tbp]
\caption{PRISM feedback loop for adaptive rate control.}
\label{alg:ctrl}
\begin{algorithmic}
\STATE Initialize $k \gets 1$
\WHILE{true}
    \STATE sleep ($\Delta$)
    \STATE $P^{\max}_{\mathrm{PRB},i}(\Delta_k) \gets \max_{j \in \mathcal{N}_{i,k}} \mathrm{PRB}_{i,j}$
    \STATE Build state $s_{i,k}$ from recent feedback observations
    \STATE Select gain $a_{i,k} \gets \pi_{\theta}(s_{i,k})$
    \IF{$P^{\max}_{\mathrm{PRB},i}(\Delta_k) < P_{\epsilon,i}$}
        \STATE $a_{i,k} \gets {a_{i,k}} / {n}$
    \ENDIF
    \STATE $r_{i,k+1} \gets r_{i,k} - a_{i,k} \cdot \gamma \cdot \left(P^{\max}_{\mathrm{PRB},i}(\Delta_k) - P_{\epsilon,i}\right)$
    \STATE $k \gets k + 1$
\ENDWHILE
\end{algorithmic}
\end{algorithm}
\subsection{PPO Formulation}
The PPO agent is implemented at the ES and selects an adaptive gain used to adjust the send rate every $\Delta$ seconds. 
Since a single feedback report only partially captures the queue and channel state, the PPO state of each UE is formed from the last eight feedback observations to approximate the Markov property~\cite{Abbasloo20}. 

Each observation contains the following normalized features: the PRB utilization (over $\delta$), minimum and maximum PRB utilization (over $\Delta)$, target PRB utilization, the TB retransmission ratio, send rate, modulation and coding scheme (MCS), and signal-to-interference-plus-noise ratio (SINR). 
\HH{MCS and SINR are complementary rather than redundant: MCS captures the realized link-adaptation outcome and saturates at high channel quality, whereas SINR reflects the instantaneous channel margin.} 
\HH{The resulting state size is $64$.} 

The PPO actor uses this state to select a discrete gain $a_{i,k}$ from the action set  $\mathcal{A}=\{0.05, 0.1, 0.2, 0.3, \ldots, 1.0\}$. 
PRISM uses two-layer multilayer perceptrons for the actor and critic, each with hidden layers of sizes $[128,64]$. 

For policy training, the reward is computed from measurements observed after each rate update.
It combines two objectives: (i) a tracking reward $\mathcal{R}_{\mathrm{track}}$ that encourages the peak PRB utilization to remain close to the target PRB utilization, and (ii) a gain-selection reward $\mathcal{R}_{\mathrm{gain}}$ that guides the policy toward an appropriate gain under different network conditions. 

To define the tracking component, we first compute the normalized error of UE $i$ during the $k$th control interval $\Delta_k$ as follows:
\begin{equation}
e_{i,k} = \frac{P^{\max}_{\mathrm{PRB},i}(\Delta_k)-P_{\epsilon,i}}{P_{\epsilon,i}}
\end{equation}
where $e_{i,k}>0$ indicates PRB overutilization and $e_{i,k}<0$ indicates underutilization. 
The tracking reward is defined as:
\begin{equation}
\mathcal{R}_{\mathrm{track},i,k} = \exp^{-1.0 \, [e_{i,k}]^-} - \exp^{2.0 \, [e_{i,k}]^+} + 1
\end{equation}
where $[x]^+=\max(x,0)$ and $[x]^-=\max(-x,0)$. 
This asymmetric form penalizes overutilization more strongly than underutilization to prevent excessive queue buildup at the BS. 

$\mathcal{R}_{\mathrm{track}}$ alone is insufficient for learning gain selection. 
When the error is small, which usually occurs under good tracking, different gain values produce similar rate updates, so an arbitrary gain within a reasonable range may perform similarly. 
However, gain selection is crucial during high-error bursts, where the tracking reward alone provides little guidance for selecting an appropriate gain. 
We therefore add a gain-selection reward that favors actions close to a target gain $a^{*}_{i,k}$: 
\begin{equation}
\mathcal{R}_{\mathrm{gain},i,k} = - \left| \frac{a_{i,k}}{\max{ (\mathcal{A}) }} - a^{*}_{i,k} \right|
\end{equation}

The target gain $a^{*}_{i,k}$ follows a fast-decrease, slow-increase strategy:
\begin{equation}
a^{*}_{i,k} =
\begin{cases}
\eta_{i,k}/\eta_{\max} & e_{i,k} > 0 \\
\frac{1}{4} \cdot \min\left(\frac{|e_{i,k}|}{0.3},1\right) \cdot {\eta_{i,k}}/{\eta_{\max}} & e_{i,k} \leq 0
\end{cases}
\end{equation}
where $\eta_{i,k}$ denotes the spectral efficiency of UE $i$ during control interval $\Delta_k$, and $\eta_{\max}$ is the maximum spectral efficiency used for normalization \cite{ETSI_138.214}. 
The factor $\eta_{i,k}/\eta_{\max}$ scales the gain with channel quality: as $\eta$ increases, each PRB carries more data, so the same PRB error requires a larger rate correction. 
When peak PRB utilization exceeds the target, $\mathcal{R}_{\mathrm{gain}}$ encourages a stronger gain, independent of the error magnitude, to sharply decrease the send rate. 
When peak PRB utilization is below the target, $\mathcal{R}_{\mathrm{gain}}$ encourages a more gradual, error-proportional rate increase scaled by the current spectral efficiency. 

To avoid large reward magnitudes and improve PPO training stability, we clip the total reward as follows: 
\begin{equation}
\mathcal{R}_{i,k} = \operatorname{clip} \left( \mathcal{R}_{\mathrm{track},i,k} + \mathcal{R}_{\mathrm{gain},i,k}, -2, 1 \right)
\end{equation}
where $\operatorname{clip} (x, \ell, u)$ constrains $x$ to the interval $[\ell, u]$. 
\HH{Since $\mathcal{R}_{\mathrm{track}} \leq 1$ and $\mathcal{R}_{\mathrm{gain}} \leq 0$, the upper bound preserves the natural maximum reward, while the lower bound prevents large tracking errors from producing excessively negative rewards that could dominate PPO updates.} 
The reward parameters were empirically tuned to penalize overutilization more strongly while maintaining stable PPO training. 
\subsection{Training and Model Selection}
\HH{PRISM performs a policy update every $512$ samples during training, with each sample collected over one control interval $\Delta$. 
It computes generalized advantage estimates (GAE) and performs $4$ PPO optimization epochs over each on-policy batch using mini-batches of $64$ samples. 
Thus, each on-policy batch is reused across $4$ epochs, with each epoch consisting of $8$ mini-batches, after which the samples are discarded and fresh on-policy experience is collected. 
Table~\ref{tab:hyper} summarizes the control and PPO hyperparameters.}

\begin{table}[tbp]
\centering
\caption{PRISM control and PPO hyperparameters.}
\label{tab:hyper}
\resizebox{1.0\columnwidth}{!}{%
\begin{tabular}{@{}llll@{}}
\toprule
Parameter & Value & Parameter & Value \\
\midrule
Feedback interval $\delta$       & $50$ms      & Discount $\gamma_{\mathrm{PPO}}$ & $0.95$ \\
Control interval $\Delta$        & $200$ms     & GAE $\lambda_{\mathrm{PPO}}$     & $0.95$ \\
RAPID baseline gain $K$          & $0.5$       & Clip $\epsilon$                  & $0.2$ \\
PRB target factor                & $0.90$      & Entropy coefficient              & $0.02$ \\
PRBs per slot                    & $106$       & Learning rate                    & $10^{-3}$ \\
Rate scaling $\gamma$            & $10^{6}$    & On-policy batch size             & $512$ \\
History length $m$               & $8$         & Mini-batch size                  & $64$ \\
State size                       & $64$        & PPO epochs                       & $4$ \\
Hidden layers                    & $[128,64]$  & $|\mathcal{A}|$                  & $11$ \\
\bottomrule
\end{tabular}
}
\end{table}

Training uses multiple $2{,}000$sec runs, each with one PRISM flow beginning at the start of the simulation and two TCP flows introduced later: one TCP Cubic and one TCP BBR. 
The TCP start times are drawn uniformly from $[300,400]$sec. Their stop times are also randomized, with the TCP Cubic flow remaining active for at least $1{,}500$sec and the TCP BBR flow for at least $750$sec, while both flows end no later than $100$sec before the end of the simulation. 
Each training run samples a new set of timings, thereby exposing the policy to diverse contention evolution patterns. 
The policy is then evaluated on a fixed $180$sec validation scenario with one PRISM flow running for the entire simulation duration and one TCP BBR flow active from $60$sec to $120$sec. 
\HH{Table~\ref{tab:scenario} summarizes the training, validation, and test scenarios.} 
\begin{table}[tbp]
\centering
\caption{Training, validation, and test scenarios.}
\label{tab:scenario}
\resizebox{0.85\columnwidth}{!}{%
\begin{tabular}{@{}lccll@{}}
\toprule
Scenario & PRISM & TCP & TCP variants & Duration \\
\midrule
Train      & $1$ & $2$ & $1$ Cubic, $1$ BBR & $2{,}000$sec \\
Validation & $1$ & $1$ & $1$ BBR            & $180$sec  \\
Test       & $2$ & $3$ & $2$ Cubic, $1$ BBR & $2{,}000$sec \\
\bottomrule
\end{tabular}
}
\end{table}

Model selection uses a validation score rather than the raw PPO reward. 
This is necessary because the reward focuses on PRB tracking and gain selection, while our evaluation objective is the goodput-latency tradeoff. 
The validation score is computed over the $60$sec overlap window and is defined as: 
\begin{equation} \label{eq:validation_score}
\mathrm{score} = \frac{T_{\mathrm{app}} \cdot (1 - P_{\mathrm{loss}})}{1 + \mathrm{owd}_{\mathrm{mean}}/5\mathrm{ms} + \mathrm{owd}_{95}/20\mathrm{ms}}
\end{equation}
where $T_{\mathrm{app}}$ is the goodput, $P_{\mathrm{loss}}$ is the packet-loss ratio, $\mathrm{owd}_{\mathrm{mean}}$ is the mean OWD, and $\mathrm{owd}_{95}$ is the $95$th-percentile OWD over the validation overlap window. 
Note that OWD measurements are used only for model selection during validation, not as PPO observation features or as inputs during test. 

We evaluate $40$ training checkpoints and select the model with the highest validation score.
\HH{Table~\ref{tab:training} shows the first $10$ checkpoints out of $40$. 
The best checkpoint occurs at checkpoint $10$: mean OWD is reduced to $3.26$ms, $p_{95}$ OWD to $4.7$ms, and maximum OWD to $11.8$ms, while the score reaches $17.74$. 
Training continues to $40$ checkpoints, but no later checkpoint improves this validation score, so checkpoint~$10$ is kept for testing.} 
\begin{table}[tbp]
\centering
\caption{Validation performance for the selected PPO training run.}
\label{tab:training}
\resizebox{1.0\columnwidth}{!}{%
\begin{tabular}{@{}clllll@{}}
\toprule
\multicolumn{1}{c}{Iteration} &
\multicolumn{1}{c}{Goodput} &
\multicolumn{1}{c}{Mean OWD} &
\multicolumn{1}{c}{$p_{95}$} &
\multicolumn{1}{c}{Max OWD} &
\multicolumn{1}{c}{Score} \\
&
\multicolumn{1}{c}{(Mbps)} &
\multicolumn{1}{c}{(ms)} &
\multicolumn{1}{c}{(ms)} &
\multicolumn{1}{c}{(ms)} &
\\
\midrule
$1$  & $37.7$ & $7.01$ & $15.7$ & $155.7$ & $11.82$ \\
$2$  & $37.7$ & $7.01$ & $15.7$ & $155.7$ & $11.82$ \\
$3$  & $36.2$ & $4.39$ & $10.9$ & $53.6$  & $14.93$ \\
$4$  & $36.0$ & $3.71$ & $7.9$  & $31.3$  & $16.82$ \\
$5$  & $35.0$ & $3.72$ & $7.6$  & $31.4$  & $16.48$ \\
$6$  & $35.3$ & $4.10$ & $8.7$  & $58.1$  & $15.64$ \\
$7$  & $33.6$ & $3.44$ & $6.0$  & $16.2$  & $16.91$ \\
$8$  & $32.4$ & $3.31$ & $5.0$  & $13.5$  & $16.96$ \\
$9$  & $32.6$ & $3.21$ & $4.6$  & $11.1$  & $17.41$ \\
$10$ & $33.5$ & $3.26$ & $4.7$  & $11.8$  & $\mathbf{17.74}$ \\
\bottomrule
\end{tabular}
}
\end{table}

\section{Evaluation} \label{sec:results}
We implement PRISM in the NS3 (v3.41) network simulator via the UDP protocol and use the 5G-LENA module~\cite{Patriciello19} to simulate non-standalone 5G networks. 
Each UE collects cross-layer measurements and reports them to the ES every $\delta = 50$ms. 
The ES aggregates the received reports, constructs the PPO state, obtains the selected gain from the PPO agent, and updates the send rate every $\Delta=200$ms. 
\HH{We use ns3-ai~\cite{Hao20} to exchange measurements and actions between NS3 and the PPO agent via a shared-memory interface.} 
\HH{Note that PRISM requires only $6$B of additional payload per feedback report compared with RAPID, which reports only PRB utilization. 
The feedback size remains constant per UE and is independent of the application data rate.} 

\HH{Next, we configure the \texttt{InH\_OfficeOpen\_LoS} channel with fast fading in the n78 band, numerology $1$, and $40$MHz bandwidth.} 
\HH{We use the PF scheduler of the 5G-LENA module, i.e., \texttt{NrMacSchedulerOfdmaPF}. 
The traffic type is configured as eMBB, and the RLC layer operates in unacknowledged mode.} 
The UEs follow the Gauss-Markov mobility model~\cite{Camp02}, with mean speed $1$m/s, memory parameter $\alpha=0.95$, and direction updates every $30$sec, within a $200{\times}200$m area centered on the BS. 
\HH{The observed MCS values are predominantly within the range $18$-$28$.} 

For both PRISM and RAPID, the target PRB utilization is selected as $P_{\epsilon,i} = 0.9 \cdot \mathrm{PRB}_{\mathrm{max},\mathrm{slot}} / n$, where $\mathrm{PRB}_{\text{max},\mathrm{slot}}$ is the maximum number of available PRBs per slot, determined by the bandwidth and the selected numerology, \HH{and $n$ is the number of active downlink flows.} 
The reduction factor of $0.9$ keeps the target below the scheduler fair-share ceiling \cite{Hassani25a} to enhance system reliability and avoid OWD spikes caused by high PRB utilization. 
\subsection{Test Results}
The test scenario includes two PRISM flows and three TCP cross-traffic flows, i.e., two TCP Cubic flows and one TCP BBR flow. 
The PRISM flows start at $1$sec and $20$sec and run until the end of the $2{,}000$sec simulation. 
The TCP flows start at $[100,200,300]$sec and stop at $[1600,1700,1800]$sec, respectively. 
Therefore, the contention level changes over time, testing whether the learned gain can react to channel variation and cross-traffic changes. 
For example, peak contention occurs from $t=300$sec, when the last TCP flow starts, to $t=1600$sec, when the first TCP flow stops. 
During this period, the target PRB utilization remains at its lowest value, $P_{\epsilon}=0.9 \cdot 106 / 5 \approx 19.1$, for more than $20$min. 
RAPID is evaluated under the same mobility, traffic timing, and TCP coexistence setup used for PRISM, with $K = 0.5$ as specified in~\cite{Hassani25a}. 

\begin{figure}[tbp]
    \centering
    \includegraphics[width=0.95\columnwidth]{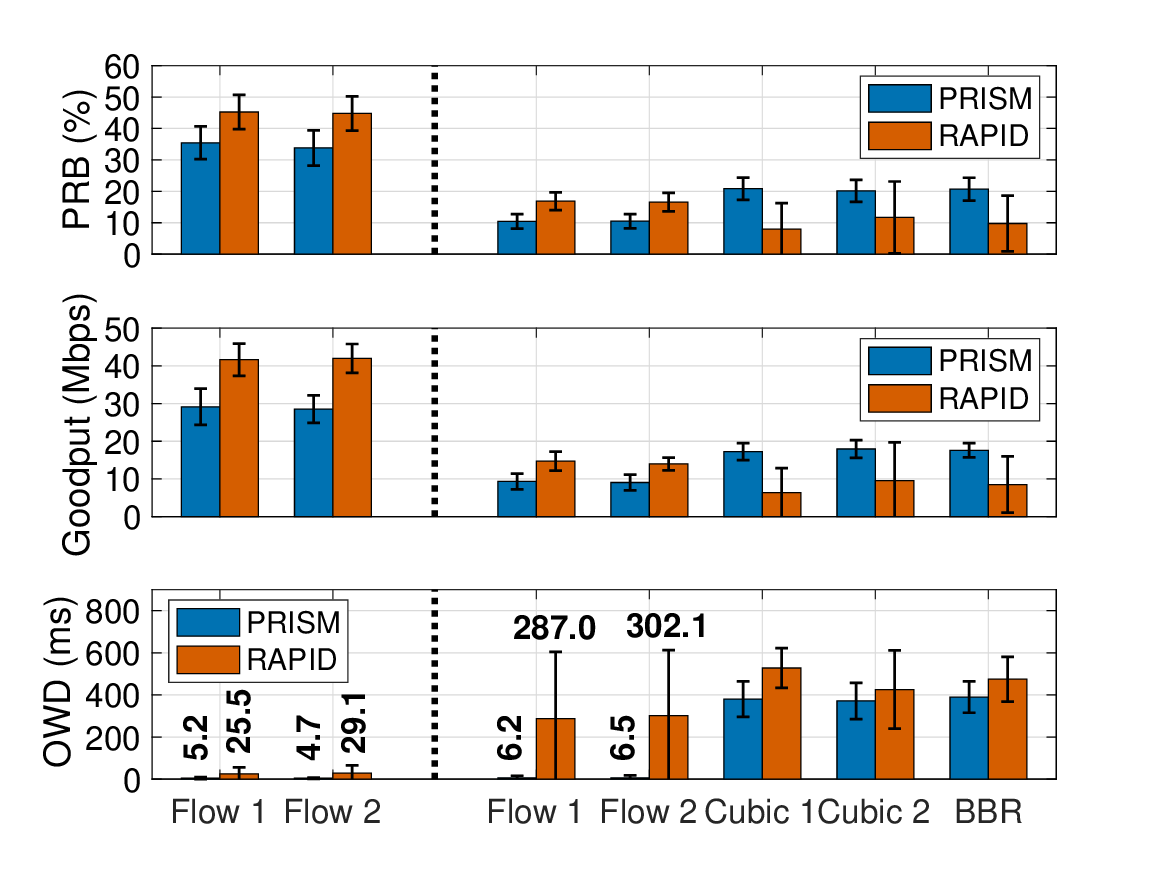}
    \caption{PRISM and RAPID performance under low and high contention during the test run. OWD values are computed on a per-packet basis. Error bars indicate one standard deviation.} 
    \label{fig:compare}
\end{figure}

Fig.~\ref{fig:compare} compares the performance of PRISM and RAPID during two intervals of the test run: a low-contention interval in which only the two PRISM/RAPID flows, i.e., Flows~1 and~2, are active, and a high-contention interval in which these flows coexist with two TCP Cubic flows and one TCP BBR flow. 
When only Flows~1 and~2 are active, PRISM maintains lower OWD for both flows, with mean OWDs of $5.2$ms and $4.7$ms, compared with $25.5$ms and $29.1$ms for RAPID, while using lower PRB utilization ($34.6\%$ vs. $45.0\%$). 
Here, $P_{\epsilon} = 47.7$ PRBs, which is $45\%$ of the total $106$ available PRBs. 

Under TCP coexistence, PRISM achieves an even larger latency reduction, with mean OWDs of $6.2$ms and $6.5$ms compared with $287.0$ms and $302.1$ms for RAPID. 
This significant reduction is achieved with lower average PRB utilization, i.e., $10.5\%$ for PRISM compared to $16.7\%$ for RAPID, against a target of $18\%$ ($P_{\epsilon} = 19.1$). 
Moreover, PRISM maintains negligible packet loss and about $9$Mbps goodput per flow, whereas the coexisting Cubic and BBR flows achieve $17.3$-$18.0$Mbps but experience substantially higher mean OWD of $371$-$390$ms. 
However, under RAPID, the TCP flows are throttled to $6.4$-$9.6$Mbps and suffer even higher OWD, ranging from $425$ms to $528$ms. 

\begin{table}[tbp]
\centering
\caption{Test results over the test run, with OWD statistics computed from per-feedback samples collected every $\delta = 50$ms.}
\label{tab:fullrun}
\resizebox{1.0\columnwidth}{!}{%
\begin{tabular}{@{}cllllll@{}}
\toprule
\multicolumn{1}{c}{Scheme} &
\multicolumn{1}{c}{Goodput} &
\multicolumn{1}{c}{Mean OWD} &
\multicolumn{1}{c}{$p_{50}$} &
\multicolumn{1}{c}{$p_{95}$} &
\multicolumn{1}{c}{$p_{99}$} &
\multicolumn{1}{c}{Pckt loss} \\
&
\multicolumn{1}{c}{(Mbps)} &
\multicolumn{1}{c}{(ms)} &
\multicolumn{1}{c}{(ms)} &
\multicolumn{1}{c}{(ms)} &
\multicolumn{1}{c}{(ms)} &
\multicolumn{1}{c}{(\%)} \\
\midrule
PRISM & $26.46$  & $5.4$   & $4.0$  & $12.1$  & $30.1$  & $0.02$ \\
RAPID & $39.35$  & $254.9$ & $15.8$ & $703.9$ & $761.5$ & $21.2$ \\
\bottomrule
\end{tabular}
}
\end{table}

Table~\ref{tab:fullrun} presents statistics for the PRISM/RAPID flows across the entire duration of the experiment. 
The results show that RAPID obtains higher goodput, but at the cost of severe queue buildup and packet loss. In contrast, PRISM backs off earlier when maximum PRB utilization exceeds the target PRB utilization, reducing mean OWD from $254.9$ms to $5.4$ms and packet loss from $21.2\%$ to $0.02\%$. The tail-latency results show the same trend: PRISM keeps $p_{99}$ OWD at $30.1$ms, while RAPID reaches $761.5$ms. 
This tradeoff is preferable for delay-sensitive interactive services. 

\begin{table}[tbp]
\centering
\caption{Test results across contention levels for PRISM and RAPID.}
\label{tab:perpe}
\resizebox{1.0\columnwidth}{!}{%
\begin{tabular}{@{}lclllllc@{}}
\toprule
& & \multicolumn{5}{c}{OWD (ms)} & Goodput \\
\cmidrule(lr){3-7}
Scheme & $P_{\epsilon}(n)$ & Mean & $p_{50}$ & $p_{95}$ & $p_{99}$ & Max & (Mbps) \\
\midrule
\midrule
PRISM & $95.4$ (1) & $3.1$ & $2.3$ & $5.6$  & $7.3$  & $8.0$   & $74.1$ \\
PRISM & $47.7$ (2) & $5.3$ & $5.3$ & $6.9$  & $11.3$ & $103.7$ & $53.3$ \\
PRISM & $31.8$ (3) & $4.0$ & $3.3$ & $5.8$  & $11.2$ & $121.6$ & $32.6$ \\
PRISM & $23.9$ (4) & $4.4$ & $3.4$ & $7.6$  & $21.9$ & $194.7$ & $23.7$ \\
PRISM & $19.1$ (5) & $5.8$ & $4.0$ & $14.9$ & $33.6$ & $169.8$ & $17.0$ \\
\midrule
RAPID & $95.4$ (1) & $4.6$   & $2.7$   & $12.9$  & $16.7$  & $18.9$  & $94.8$ \\
RAPID & $47.7$ (2) & $27.5$  & $10.1$  & $114.2$ & $160.2$ & $253.4$ & $77.2$ \\
RAPID & $31.8$ (3) & $181.5$ & $30.4$  & $424.4$ & $464.2$ & $647.9$ & $46.2$ \\
RAPID & $23.9$ (4) & $236.3$ & $42.3$  & $550.7$ & $600.0$ & $699.8$ & $34.9$ \\
RAPID & $19.1$ (5) & $319.4$ & $109.8$ & $720.5$ & $773.6$ & $973.5$ & $27.1$ \\
\bottomrule
\end{tabular}
}
\end{table}

\HH{We break down the same test run by contention regime in Table~\ref{tab:perpe}. 
Each regime corresponds to a different target PRB utilization $P_{\epsilon}$, which changes as flows start and stop during the run. 
The OWD statistics in Table~\ref{tab:perpe} are computed from the per-feedback samples, each representing the mean OWD over a $\delta = 50$ms reporting interval.} 

\HH{The contention-regime breakdown further shows that PRISM keeps $p_{99}$ OWD within tens of milliseconds across all regimes, whereas RAPID degrades sharply as contention increases. 
This disparity is most pronounced at the lowest target PRB utilization, indicating that PRISM is most valuable under high contention. 
When $P_{\epsilon}=19.1$, PRISM reduces mean OWD from $319.4$ms to $5.8$ms ($55\times$ reduction) and reduces $p_{99}$ OWD from $773.6$ms to $33.6$ms ($23\times$ reduction). 
This latency improvement comes at the cost of a $37.3\%$ decrease in goodput ($37.4\%$ reduction in PRB utilization) compared with RAPID,}
\IN{indicating that PRISM trades bandwidth utilization for lower delay by backing off more aggressively to prevent excessive RLC queue buildup.} 

\IN{To verify that RAPID's degradation is not primarily caused by the choice of its fixed feedback gain, we repeat the $2{,}000$sec test for $K \in \{0.1, 0.25, 0.5, 1, 2, 5\}$. 
Across the tested values, RAPID's mean OWD remains between $255$ and $497$ms, $p_{99}$ OWD between $760$ and $787$ms, and packet loss between $21\%$ and $36\%$. 
Notably, $K = 0.5$ achieves the highest goodput and the lowest mean OWD and packet loss. 
For all tested gain values, average PRB utilization remains $7$-$14\%$ below $P_{\epsilon}$, while peak PRB utilization exceeds $P_{\epsilon}$ by $17$-$20\%$, indicating that tuning $K$ alone does not eliminate the high-delay behavior of RAPID.}

\subsection{PPO Gain Analysis}
\HH{We analyze the PPO-selected gain $a_{i,k}$ to better understand how PRISM reduces tail latency. We aggregate the control decisions of the two PRISM flows over the full test run, since both use the same policy.} 
The selected gains are concentrated near the lower end of the action space $\mathcal{A}$: gain $0.05$ accounts for about $75\%$ of all decisions and gain $0.2$ for about $18\%$, while larger gains in the range $0.4$-$0.5$ appear in only about $5\%$ of cases. 
\begin{figure}[tbp]
    \centering
    \includegraphics[width=0.72\columnwidth]{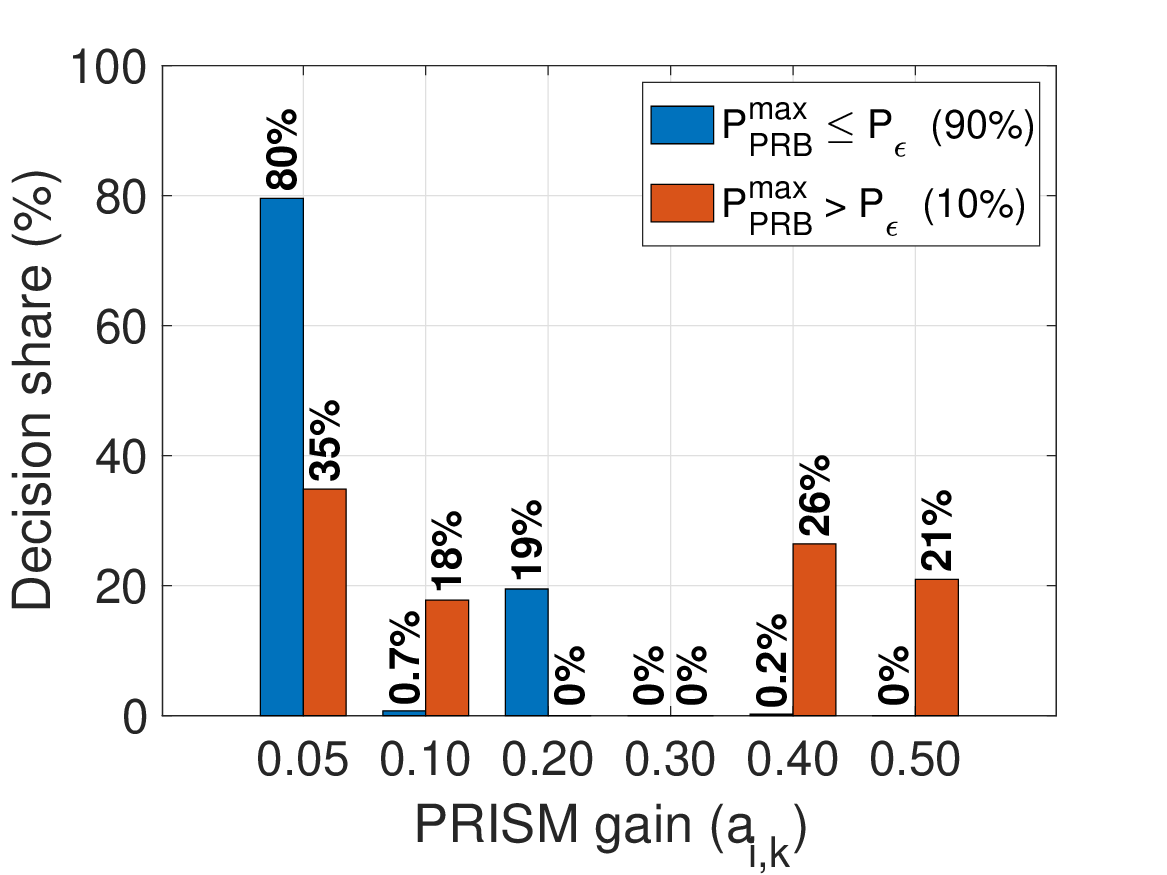}
    \caption{Distribution of the PPO-selected gain $a_{i,k}$ over the test run. The policy does not select gains in the range $0.6$-$1.0$. NS3 simulation results.} 
    \label{fig:gain}
\end{figure}

Fig.~\ref{fig:gain} shows this behavior more clearly by splitting the selected gains based on whether the peak PRB utilization falls below or exceeds the target PRB utilization $P_{\epsilon}$. 
When $P^{\max}_{\mathrm{PRB}} \leq P_{\epsilon}$, which represents roughly $90\%$ of the total decisions, the policy selects the minimum gain of $0.05$ in about $80\%$ of the cases and only infrequently uses gains above $0.2$. 
However, when overshoot occurs, the policy shifts toward higher gains, with $0.4$ and $0.5$ selected in approximately $47\%$ of such instances. 
Thus, PRISM induces only small rate adjustments during typical operation but responds more aggressively when the peak PRB utilization surpasses the target PRB utilization and queue buildup is likely. 
This selective use of larger gains is consistent with the tail-latency reductions in Table~\ref{tab:fullrun}. 
\subsection{Impact of the Design Components}
\IN{The performance gains of PRISM result from complementary design choices: peak PRB utilization captures short resource bursts that may precede queue buildup, while the adaptive PPO gain controls the magnitude of rate adjustments under changing conditions. 
To assess the contribution of each component, we repeat the test run with an intermediate configuration that combines the peak statistic with the fixed gain $K = 0.5$. 
Table~\ref{tab:ablation} compares this configuration with RAPID and PRISM.} 

\begin{table}[tbp]
\centering
\caption{Impact of the two design components of PRISM over the $2{,}000$sec test run.}
\label{tab:ablation}
\resizebox{1.0\columnwidth}{!}{%
\begin{tabular}{@{}llllllll@{}}
\toprule
\multicolumn{1}{c}{Scheme} &
\multicolumn{1}{c}{PRB} &
\multicolumn{1}{c}{Feedback} &
\multicolumn{1}{c}{Goodput} &
\multicolumn{1}{c}{Mean OWD} &
\multicolumn{1}{c}{$p_{95}$} &
\multicolumn{1}{c}{$p_{99}$} &
\multicolumn{1}{c}{Pckt loss} \\
&
\multicolumn{1}{c}{statistic} &
\multicolumn{1}{c}{Gain} &
\multicolumn{1}{c}{(Mbps)} &
\multicolumn{1}{c}{(ms)} &
\multicolumn{1}{c}{(ms)} &
\multicolumn{1}{c}{(ms)} &
\multicolumn{1}{c}{(\%)} \\
\midrule
RAPID        & Mean & $0.5$ & $39.35$ & $254.9$ & $703.9$ & $761.5$ & $21.2$ \\
Intermediate & Peak & $0.5$ & $34.34$ & $15.7$  & $46.8$  & $405.0$ & $0.23$  \\
PRISM        & Peak & PPO   & $26.46$ & $5.4$   & $12.1$  & $30.1$  & $0.02$  \\
\bottomrule
\end{tabular}
}
\end{table}

\IN{The intermediate configuration substantially lowers mean OWD from $254.9$ms to $15.7$ms and packet loss from $21.2\%$ to $0.23\%$, with a $12.7\%$ reduction in goodput. 
However, $p_{99}$ OWD remains at $405.0$ms. 
With the PPO-selected adaptive gain, PRISM further reduces mean OWD to $5.4$ms and $p_{99}$ OWD to $30.1$ms, at an additional $22.9\%$ reduction in goodput. 
These results show that the peak PRB statistic primarily reduces mean OWD and packet loss, while the adaptive gain provides additional control over tail latency.} 
\subsection{Robustness Across Test Seeds}
\HH{To assess robustness, we evaluate the PRISM policy over $10$ independent test seeds using the entire $2{,}000$sec test scenario, varying the fading and mobility realizations while keeping the traffic timing fixed. 
Across the $10$ seeds, the median $p_{99}$ OWD is $27.5$ms for PRISM~1 and $30.7$ms for PRISM~2, with corresponding standard deviations of $12.1$ms and $11.4$ms. 
The observed $p_{99}$ OWD ranges from $10.8$-$51.8$ms across both PRISM flows. 
These results indicate that PRISM maintains low tail latency despite variation in wireless conditions.} 
\section{Conclusion} \label{sec:conclusion}
This paper introduced PRISM, a PPO-based CCA for mobile networks. 
PRISM leverages cross-layer metrics, including PRB utilization statistics, MCS, and TB retransmission ratio, to regulate the send rate. 
The NS3 simulation results show that PRISM significantly reduces OWD under mobility and TCP cross-traffic, achieving a favorable tradeoff between latency and goodput for delay-sensitive applications. 
\HH{The multi-seed evaluation further demonstrates that PRISM maintains low tail latency across varying fading and mobility settings, highlighting it as a promising low-latency CCA for mobile networks.} 
We plan to benchmark PRISM against other RL-based CCAs and analyze its sensitivity to design parameters, as well as its robustness, fairness, and scalability. 
Moreover, a key objective of our future work is to implement PRISM in a 5G testbed and evaluate its performance under real radio conditions, \HH{including its decision-making latency and computational overhead for real-time applications.} 
\IN{Future work may explore distinct CCAs for different application classes, e.g., delay-critical and bandwidth-oriented traffic, by varying the PRB statistic used for rate control and the reward weighting.} 
\section*{Acknowledgments}
\IN{This work was supported by the HICONNECTS project under Grant Agreement No.~101097296 and by the Dutch National Growth Fund program Luchtvaart in Transitie (LiT).} 
\bibliographystyle{IEEEtran}
\bibliography{references}

\end{document}